# An empirical study of large, naturally occurring starling flocks: a benchmark in collective animal behaviour

MICHELE BALLERINI[1,2], NICOLA CABIBBO[3,4], RAPHAEL CANDELIER[3], ANDREA CAVAGNA[1,5], EVARISTO CISBANI[2], IRENE GIARDINA[1,5], ALBERTO ORLANDI[1], GIORGIO PARISI[1,3], ANDREA PROCACCINI[1,3], MASSIMILIANO VIALE[3] & VLADIMIR ZDRAVKOVIC[1]

[1]*Centre for Statistical Mechanics and Complexity (SMC), CNR-INFM;* [2]*Istituto Superiore di Sanita'(ISS);* [3] *Dipartimento di Fisica, Universita' di Roma 'La Sapienza';* [4]*Istituto Nazionale di Fisica Nucleare,* [5]*Istituto dei Sistemi Complessi (ISC), CNR;*

**Correspondence to:**

Irene Giardina

SMC, CNR-INFM, Dipartimento di Fisica, Universita' di Roma 'La Sapienza', P.le Aldo Moro 2, 00185 Roma, Italy

irene.giardina@roma1.infn.it

M. Ballerini, A. Cavagna, A. Orlandi, A. Procaccini, V. Zdravkovic:

SMC, CNR-INFM, Dipartimento di Fisica, Universita' di Roma 'La Sapienza', P.le Aldo Moro 2, 00185 Roma, Italy

N. Cabibbo, G. Parisi:

Dipartimento di Fisica, Universita' di Roma 'La Sapienza', P.le Aldo Moro 2, 00185 Roma, Italy

E. Cisbani:

ISS, viale Regina Elena 299, 00161 Roma, Italy

M. Viale (current address):

Dipartimento di Fisica, Universita' di Roma 3, via della Vasca Navale 84, 00146 Roma, Italy

R. Candelier (current address):

GIT / SPEC / DRECAM, Bat. 772, Orme des Merisiers, CEA Saclay, 91191 Gif sur Yvette, France

**Abstract:**

Bird flocking is a striking example of collective animal behaviour. A vivid illustration of this phenomenon is provided by the aerial display of vast flocks of starlings gathering at dusk over the roost and swirling with extraordinary spatial coherence. Both the evolutionary justification and the mechanistic laws of flocking are poorly understood, arguably because of a lack of data on large flocks. Here, we report a quantitative study of aerial display. We measured the individual three-dimensional positions in compact flocks of up to 2700 birds. We investigated the main features of the flock as a whole - shape, movement, density and structure - and discuss these as emergent attributes of the grouping phenomenon. We find that flocks are relatively thin, with variable sizes, but constant proportions. They tend to slide parallel to the ground and, during turns, their orientation changes with respect to the direction of motion. Individual birds keep a minimum distance from each other that is comparable to their wingspan. The density within the aggregations is non-homogeneous, as birds are packed more tightly at the border compared to the centre of the flock. These results constitute the first set of large-scale data on three-dimensional animal aggregations. Current models and theories of collective animal behaviour can now be tested against these results.

The aerial display of large flocks of birds is a stunning example of collective behaviour in animal aggregations (Emlen 1952). A paradigmatic case is provided by European starlings (*Sturnus vulgaris*) (Feare 1984). These birds can be observed in many cities, where they establish their roosting sites. Shortly before sunset starlings return to their roost and, prior to retiring for the night, they form sharp-bordered flocks, ranging from a few hundred to tens of thousands of birds, which wheel and turn over the roosting site until darkness falls. Flocks exhibit strong spatial coherence and are capable of very fast, highly synchronized manoeuvres, either spontaneously, or as a response to predator attacks. Many features of bird flocking are present in other instances of collective animal behaviour. Fish schools, mammal herds and insect swarms represent other examples of animal aggregations that have fascinated biologists for many years (Gueron et al. 1996; Parrish & Edelstein-Keshet 1999; Krause & Ruxton 2002; Couzin & Krause 2003). Like starlings, individuals form cohesive groups that are able to sustain remarkable coordination.

Diverse instances of collective behaviour are found in many different fields of science, from the spontaneous ordering of magnetic moments in physics (see, e.g., Cardy 1996), the coordination of an ensemble of artificial agents with distributed intelligence in robotics(Cao et al. 1997; Jadbabaie et al. 2003), the emergence of herding behaviour in financial markets in economics (Cont & Bouchaud 2000), to the synchronized clapping in a concert hall (Neda et al. 2000; Michard & Bouchaud 2005) or the Mexican wave in a stadium (Farkas et al. 2002). In all these examples, collective behaviour emerges as the result of the local interactions between the individual units, without the need for centralized coordination. The tendency of each agent to imitate its neighbours (*allelomimesis*), can, by itself, produce a global collective state. Whenever this happens, we are in the presence of *self-organized* collective behaviour.

Although self-organization is undoubtedly a general and robust mechanism, its universality is an open issue. In physics, for example, universality is a well-defined concept: the same model and theory can be used to describe quantitatively very different physical systems, provided that they all share the same fundamental symmetries. The situation is more complicated in biology because the individuals that form a group are much more complex than particles or spins. For example, despite the fact that fish schools and bird flocks behave similarly, certain collective patterns are present in one case and not in the other (Krause & Ruxton 2002). At some level, the specificities of the individuals and of the environment must make a difference. Therefore, in view of the highly interdisciplinary nature of self-organized collective behaviour, it is important to distinguish the general from the specific.

Models play a crucial role in this respect. Indeed, it was modelling exercises that revealed the general principles of how collective behaviour can emerge from self-organization. When it comes to modelling real instances of collective behaviour however we need to be more detailed. In this case, models must specify the minimal conditions necessary to reproduce the empirical observations, so that we can distinguish between general phenomena and those specific to the system.

The field of collective animal behaviour boasts a wealth of models (Aoki 1982; Heppner & Grenader 1990, Reynolds 1987; Huth & Wissel 1992; Couzin et al. 2002; Inada & Kawaki 2002, Kunz & Hemelrijk 2003; Vicsek et al. 1995; Gregoire & Chaté 2004). Some of these were developed for fish schools, some for bird flocks, and some with a non-specific biological target. In all cases, however, the models agree on three general behavioural rules: move in the same direction as your neighbours; remain close to them; avoid collisions. These rules are modelled using three distinct contributions to the interaction among the individuals, i.e. alignment of velocities, attraction and short-range repulsion. In all cases, the models produce cohesive aggregations that look *qualitatively* similar to natural cases. However, each model has its own way of implementing the rules, dictated by the differing opinions as to which are the relevant conditions, and by the different biological targets (e.g. fish vs. birds). Of course, selection among different models can only be achieved by comparing their results with empirical evidence. Only empirical observations can tell us whether or not the collective properties of a model are in *quantitative* agreement with the natural case. Moreover, the feedback between models and empirical data must confirm whether or not a certain condition is truly necessary to reproduce a specific biological feature.

Empirical data, then, are necessary both as a crucial input of the modelling approach, and as a quantitative guideline for answering more fundamental questions about groups, their global features and biological function (see e.g. Parrish & Hammer 1997). Unfortunately, 3D data on even moderately large groups of animals are very hard to obtain, and quantitative empirical data are scarce and limited to very small groups (a few tens of individuals). Testing of the models has therefore been sporadic so far. At the same time, speculation on the microscopic origin and biological function of collective behaviour has outgrown empirical groundwork.

Empirical 3D studies on fishes have been performed in laboratory tanks (Cullen et al. 1965; Partridge et al. 1980; Partridge 1980; Van Long et al. 1985; Tien at al. 2004). Data for birds, in contrast, have been obtained in field observations (Miller and Stephen 1966; van Tets 1966; Major & Dill 1978; Pomeroy & Heppner 1992; Budgey 1998). These past studies have however two major limitations: the number of individuals is small (limited to few tens) and the group arrangements were loose, at variance with the huge, highly cohesive groups characteristic of collective behaviour. Both these drawbacks stem from a single technical problem: in order to reconstruct the 3D position of an object, all optical techniques (stereometry, orthogonal method, shadow method) require different images to be placed in correspondence (i.e. to be matched, see Fig.1a,b) (Osborn 1997; Hartley & Zisserman 2003). For large and compact sets of featureless points, this problem is so severe that it has been suggested that these techniques are fundamentally inadequate to handle 3D biological scenes (Aloimonos et al. 1991).

Using statistical physics, optimization theory and computer vision techniques we have managed to solve the correspondence problem. We developed an experimental technique capable of reconstructing the individual 3D positions in cohesive aggregations of several thousands of animals in the field (see Cavagna et al. 2008a). We used this technique to collect quantitative empirical data on large flocks of starlings during aerial display. In this paper, we present quantitative and systematic data on the two main attributes of the groups: global properties (shape, size, orientation and movement) and internal structure (density profile and neighbours distributions). Our aim in doing so is two-fold:

**i.** We wish to provide a detailed analysis of the mechanistic laws of flocking, at the global and structural level. This enables us to set a new empirical benchmark for testing existing models of self-organized collective behaviour.

**ii.** We wish to characterize the attributes of flocks as emergent properties of the grouping phenomenon. To this end, we attempt to place our results in the context of the biological function of grouping, individual fitness consequences, interaction with the environment and mutual interaction between individuals.

## METHODS

Both the experimental technique employed to obtain 3D data and the statistical tools used to analyse the data are reported in detail elsewhere (Cavagna et al. 2008 a,b). In the present section, therefore, we give only a brief summary of the most essential points.

### Location and Materials

Large colonies of European Starlings (*Sturnus vulgaris)* spend the winter in Rome, Italy, where several roosting sites are located within the city urban area. In our analysis we do not consider migrating flocks, or travelling flocks, i.e. flocks that move from feeding to roosting site. Instead, we focus only on those engaged in aerial display: shortly before dusk, very cohesive flocks of starlings swirl over the roost wandering in a wide but confined area. We took digital images of starling aerial display at the roosting site of Termini railway station, between December 2005 and February 2006, from the terrace of Palazzo Massimo, Museo Nazionale Romano. The apparatus was located 30m above ground level. Wind speed never exceeded 12ms$^{-1}$. Average distance of birds from the cameras was 100m. We used Canon EOS 1D Mark II digital cameras (3504×2336 pixels), mounting Canon 35-mm lenses. The aperture was set between f2.0 and f4.0; shutter speed between 1/1000s and 1/250s; ISO between 100 and 800; and cameras' tilt-up between 35% and 40%.

### Experimental technique.

We used stereo photography to collect data on flocks (Longuet-Higgins 1981; Osborn 1997). More specifically, we employed the trifocal technique (Hartley & Zisserman 2003), with three different points of view. The distance between the two stereo cameras (the baseline) was $d$=25m. Given two targets at 100m from the apparatus, with a relative distance of 1m, our tests gave an error of $\delta z$<0.04m on the relative distance, and an error of $\Delta z$<0.92m on the absolute distance of the targets. The error is thus below 5% for both the relative and absolute distance. Our apparatus shot series of photographs at a rate of 10fps (frames-per-seconds), for a maximum of 8 seconds (80 consecutive photographs). After that, the cameras' buffer filled up and we had to wait a few seconds before another series of photographs could be taken. If the flock under consideration exited from the field of view of any of the cameras, the series was interrupted. The algorithms used to solve the correspondence problem and perform the 3D reconstruction are described in Cavagna et al. (2008a). In the cases we analysed, we reconstructed on average 88% of the birds, and never less than 80% of the flock.

### Events selection

A flocking event was defined as a series of up of 80 consecutive photographs shot at 10fps of a single flock. We collected on average 15 events every session, and we thus gathered approximately 500 flocking events over 40 sessions. Not all of these events, however, could be processed. The vast majority of them were not in the common field of view of all cameras, a necessary condition to perform the 3D reconstruction. In addition, the distance of the flock had to be smaller than 250m (due to our photographic resolution); the total number of objects in the photograph had to be smaller than 8000 birds (due to software constraints); and exposure and contrast had to be appropriate. Only 50 events met these criteria. We then selected 20 of these, choosing flocks with sharp borders, strong spatial cohesion and a large number of birds ($N > 400$). Finally, our matching algorithm put a limit to the flock's maximum density on the photographs, leaving only 10 events suitable for analysis. Using synthetic data, we checked that the reconstruction software did not introduce any significant bias in the flocks' shape and structure.

### Biological target of investigation

Qualitatively, we could distinguish between two classes of aerial display. Some flocks flew very high above the roost (>200m), were typically very large (tens of thousands of birds) with a columnar-like shape and performed the most striking changes in formation patterns. These flocks were visible from very far away, due to their size, altitude and change in density. It has been suggested that these flocks may act as beacon to signal the location of the roost to co specifics (Feare 1984). These groups were too far away to allow us to use our reconstruction procedure effectively, as optical resolution is the main bottleneck. Moreover, the pattern of these flocks seems to be specific to starlings, whereas we were interested in more general features of collective behaviour. Therefore, we do not address these kinds of flocks here. The other kind of flock flew at a lower altitude (<100m), right over the roost. They were usually smaller in size, ranging from a few hundred to several thousand birds and were often very compact, with sharp-bordered edges. These flocks performed a random walk above the roost, keeping the same global features for quite long time scales. In other words, they exhibited a form of steady-state behaviour, which made them a good target for statistical investigation. These flocks are the ones we were able to reconstruct accurately and the present study therefore focuses on the properties of these flocks.

### Methods of analysis

The reconstruction provided us with the 3D coordinates of each individual bird in the flock. The next step was to identify the border of the flock. Standard tools, such as the convex hull, are inadequate given the presence of concavities at different scales. Rather, we used the α-shape algorithm (Edelsbrunner & Mucke 1994), a method that allows the pinpointing of concavities of the boundary down to a given scale α. This algorithm identifies a non-convex border that appropriately follows the external concavities and faithfully reproduces the external shape. This procedure allowed us to identify birds that belonged to the border and all the others were therefore identified as 'internal' birds. Typically, birds on the border, or close to it, have statistical properties that are different from internal birds. When computing structural features (like neighbour distribution, average distances and so on) one needs to take into account these boundary effects in order to produce unbiased data. In all our analysis, we used appropriate techniques to deal with this problem, as explained in Cavagna et al. (2008b).

Once the border was defined, the volume of the flock was computed via the Delaunay triangulation, identifying the sub-ensemble of the triangulation internal to the border. Flocks typically have asymmetric shapes: they are thinner in one direction and more extended along the other two. To characterize this shape quantitatively, we defined the three main dimensions and axes of the flock (see Cavagna et al. 2008b). The shortest dimension, the thickness $I_1$, was defined as the diameter of the largest sphere contained within the flock's boundary. We then fitted a plane to the flock and defined the axis orthogonal to this plane as the *yaw* axis relative to $I_1$. We refer to this axis by means of a unit vector $\mathbf{I_1}$ parallel to it. Since the flock extends in space mostly orthogonally to $\mathbf{I_1}$, we could exploit the fitted plane to identify the two largest dimensions. The flock was then projected onto this plane. Following a similar strategy as before, we defined the second dimension $I_2$ as the diameter of the smallest circle contained in flock's projection on the plane. We then fitted the 2D projection of the flock to a line, which identified the axis $\mathbf{I_3}$ along the flock and the relative largest dimension $I_3$. The orthogonal direction to $\mathbf{I_3}$ on the plane defined the axis $\mathbf{I_2}$ relative to the intermediate dimension $I_2$. The three unitary axes $\mathbf{I_1}$, $\mathbf{I_2}$ and $\mathbf{I_3}$ are by construction orthogonal to each other and can be used to describe the orientation of the flock in space. For example, the main axis $\mathbf{I_3}$ represents the direction along which the flock is more extended: by measuring the mutual orientation of $\mathbf{I_3}$ with the velocity we can identify whether the flock is elongated along the direction of motion.

The three dimensions $I_1$, $I_2$ and $I_3$ give a quantitative description of the flock shape. For example, if $I_1$, $I_2$ and $I_3$ assume values comparable one to the other this means that the flock has an approximately symmetrical shape (in the case of a sphere or of a cube, one finds $I_1 = I_2 = I_3$). If $I_1$ is much smaller than $I_2$ and $I_3$ then the group is very thin along the smallest axes and extends more perpendicularly to it. From these examples it is clear that what matters are the relative and not the absolute values of these dimensions. For this reason we considered ratios between the dimensions in our analysis (see Results).

# RESULTS

*Morphology*

A typical example of a flock, together with its 3D reconstruction, is shown in Fig.1 a-f. By rotating the point of view, the striking thinness of the flock is immediately apparent (Fig. 1c). To characterize the shape and thinness of the flocks, we consider its three dimensions $I_1<I_2<I_3$ and measure the aspect ratios, i.e. the ratios of the two longer dimensions to the shortest one: $I_2/I_1$ and $I_3/I_1$ (Table 1). Despite some fluctuations across flocks, these ratios are generally stable, and do not show any significant dependence on the number of birds or the volume of the flocks (see Fig. 2b). The average over all events gives: $I_2/I_1=2.8\pm0.4$ and $I_3/I_1=5.6\pm1.0$ (confidence intervals of 95% throughout the paper), confirming the visual impression that flocks are generally thin.

In contrast, the flocks' thickness $I_1$, i.e. the shortest dimension is highly variable. We can explain this by noting that, if the aspect ratios $I_2/I_1$ and $I_3/I_1$ are constant, the thickness $I_1$ must be linearly correlated to $V^{1/3}$, where $V$ is the flock's volume. This is accurately verified by our data, as shown in Fig. 2a (Pearson correlation test: $R^2=0.97$, $n=10$, p < 0.001). Therefore, while the number of individuals and flock volume changes significantly across flocks, flocks proportion remains approximately constant.

The concavity of a flock can be quantified as the relative difference between the volume of the convex hull (smallest convex polyhedron enclosing the flock) and the volume of the aggregation computed with the α-shape method. The results are reported in Table 1. Some flocks do exhibit a nonzero concavity, which indicates that defining their border with the convex hull would be inappropriate, since this method would include large portions of space devoid of animals. Unlike the aspect ratio, however, there is no typical value for the concavity.

*Orientation*

Given the highly non-spherical shape of flocks, we can ask whether or not they have a preferential orientation in space. To investigate this, we looked at the flock axes $\mathbf{I_1}$, $\mathbf{I_2}$, and $\mathbf{I_3}$. We define the unit vector $\mathbf{G}$ as parallel to gravity and compute the inner product of $\mathbf{I_1}$ and $\mathbf{G}$, i.e. the cosine between yaw and gravity. This quantity is close to 1 in most flocks (Table 1 and Fig. 3) and, on average, we find $|\mathbf{I_1}\cdot\mathbf{G}|=0.93\pm0.04$. This shows that the yaw axis is approximately parallel to gravity, and therefore that the flock's plane is parallel to the ground. It is also interesting to study the relationship between orientation in space and global movement of the flock. We define the centre of mass as the average position of all birds in the flock, and then measure its direction of motion. Such direction is identified by a unit vector $\mathbf{V}$, representing the normalized velocity. The orientation of flight can be described by the scalar product of the unitary velocity $\mathbf{V}$ with gravity and yaw (Table 1). Both these quantities are small for most events (see Fig. 3). We find, on average, $|\mathbf{V}\cdot\mathbf{G}|=0.13\pm0.05$ and $|\mathbf{V}\cdot\mathbf{I_1}|=0.19\pm0.08$. Velocity is thus approximately perpendicular to gravity and yaw. This result, together with the fact that yaw and gravity are nearly parallel to each other, implies that most of the time flocks slide horizontally, flying approximately parallel to the ground.

Finally, we can look at the orientation of the flock with respect to the other two axes, $\mathbf{I_2}$ and $\mathbf{I_3}$. The scalar product of the velocity with each one of these two vectors changes significantly from flock to flock, and, in some cases, it also changes across time. Some flocks appear to be elongated in the direction of motion (i.e. the velocity direction), having a large value of $|\mathbf{V}\cdot\mathbf{I_3}|$, whereas for other flocks it is the opposite, with a large value of |**V·I₂**|. We do not find, therefore, any correlation between direction of motion and elongation of the aggregation.

*Turns*

An interesting problem to consider is how birds coordinate their movements during turns. This requires analysis of events that are long enough to display a turn. Event 32-06 is almost 4 seconds long, it moves parallel to the ground (Fig.4b) and the projection of its trajectory on the plane perpendicular to gravity (the horizontal plane) shows a clear turn (Fig. 4a). The unit vectors **I₂** and **I₃** associated to the dimensions $I_2$ and $I_3$ identify the two orthogonal axis of the flock on the horizontal plane. The aspect ratio of this flock on the plane remains approximately constant throughout the turn ($I_3/I_2$ ~ 4/3). We find that the turn is accompanied by a rotation of the velocity **V** with respect to **I₂** and **I₃**. This can be seen clearly from the time evolution of the angles between the projection of the velocity **V** on the plane, and the axes **I₂** and **I₃**: for example, the angle between the projection of **V** and **I₃** is slightly larger than 90 degrees at the beginning of the curve, whereas it is 0 degrees at the end of the turn (Fig. 4c). Hence the flock's orientation changes with respect to the velocity, but it remains approximately constant with respect to an absolute reference frame (Fig. 4d). We find the same in three other flocks that displayed turns (events 17-06, 25-08, 28-10).

Another interesting behaviour can be observed in event 32-06. Before the turn the flock performs standard level flight, sliding horizontally. When the turn begins, however, the flock tilts the yaw axis **I₁** with respect to the vertical, and thus the angle between **I₁** and **G** increases (Fig. 4e). At the same time, **I₁** acquires a nonzero component in the direction of motion, and therefore the angle between **I₁** and **V** decreases. These two results indicate that when the turn is initiated the yaw axis of the flock rolls (or banks) in the direction of the turn (velocity and gravity remain orthogonal to each other). As a consequence, during a turn, the flock's plane does not slide parallel to the ground, but undergoes a finite drag. This manoeuvre is reminiscent of an aircraft's banking turn. However, the aircraft rotates in the direction of the turn (the front of an aeroplane always remains the same) and the mutual orientation of the velocity and the main aircraft axes (**I₂** and **I₃**) does not change. In the case of a flock, as we have seen, the velocity rotates with respect to the main axes (what was the front becomes the side and vice versa). Moreover, there is no global lift on the flock, as it is not a unique rigid body. Therefore, even though individual birds will bank during the turn, it is unclear why flock as a whole also does so, rather than simply sliding. A possible explanation is that it allows birds to keep their neighbours on the same visual plane, preserving visual information on their neighbours' positions.

*Average Density and Nearest Neighbour Distance*

When looking at flocks, one of the first questions concerns the density, or the degree of compactness of the aggregations. We estimate the density ρ as the number of internal birds divided by the volume of flock, as defined with the α-shape (see Methods). The values obtained for the ten flocking events are reported in Table 1 and show that density varies substantially across flocks. We find that ρ does not depend on the number of birds belonging to the flock, nor on the velocity (P =0.39 and 0.16 respectively). An alternative measure of compactness (or rather of sparseness) is given by the average nearest neighbour distance $r_1$. We find that $r_1$ is sharply related to the density, ρ being proportional to $r_1^{-3}$, as expected for an homogeneous arrangement of points (Stoyan & Stoyan 1994) (Fig.5a). The nearest neighbour distance $r_1$ also does not depend on the size of the group, contrary to the pattern observed in fish schools (Partridge at al. 1980) and in computer simulations (Kunz & Hemelrijk 2003). The values of $r_1$ are shown in Figure 5b and once again reveal that the flocks we analysed were quite heterogeneous: nearest neighbour distance varied more than 100% between densest and sparsest events.

### Density Variations Within the Aggregation

A further important question concerns density variations through the group: is there any significant density gradient in the flock? To answer this question we investigated how density or, alternatively, sparseness, changed as one moved from the border to the centre of the flock. We divided the flock into shells of thickness δ, moving from the border to the centre. One shell is defined as the subset of birds that are found at a distance $d < k\,\delta$ from the border, with $k = 1, 2, 3…$. Within each shell, we measured the average nearest neighbour distance. In this way, we obtained the nearest neighbour distance $r_1$ as function of the distance from the border, $d$. The behaviour of $r_1(d)$ for our flocks is radically different from the one of a homogeneous system (Fig. 6a). It shows that the nearest neighbour distance increased from the border to the centre and indicates that flocks had a higher density close to the border (small $r_1$) and were more sparse in the centre (large $r_1$). We also looked directly at a density measure by defining a gradient function from the border to the centre. For a given value of δ, we erased all the birds with $d < \delta$, i.e. the birds close to the original border. We were then left with a reduced set of birds where the outer shell had disappeared. We computed the border of this new set using the α-shape method, its volume and, finally, its density. We iteratively re-computed the density until we reached the core of the flock. For δ=0 this basically corresponds to progressively peeling away layers of the flock. As can be seen in Fig. 6b, this measure shows that density is higher close to the border. This density gradient is larger in some flocks than in others but it is qualitatively present in all 10 flocks we analysed. To investigate whether there is also a front to back density imbalance, we computed the balance shift, defined as the relative difference along the direction of motion between the position of the centre of mass and the geometrical centre of the flock. Positive values of the balance shift mean that the centre of mass is located more toward the front of the flock with respect to the geometrical centre, and that there is therefore a neat imbalance of birds in the front. The values of the balance shift for all the analysed events are reported in Table 1. Sometimes the birds were concentrated more toward the front of the flock (4 flocks), sometimes more toward the back (2 flocks), and sometimes they were uniformly distributed in the outer shell.

### Nearest Neighbours Radial Distribution and Exclusion Zone

To gain an insight into the inter-individual structure of the flock, we looked at the probability distribution of the nearest neighbour (n.n.) distance, $P(r)$ (Fig. 7). The shape of this function is reminiscent of a random (Poisson) set of points. There are, however, important differences. Most notably we observed that $P(r)$ displayed a drop at low values of $r$ and, correspondingly, that the cumulative probability $P > (r)$ displayed a shift compared to the Poisson case. The biological origin of this drop is rather intuitive: birds maintained a certain distance from each other in order to avoid collisions. For this reason, most numerical models of collective behaviour assume the existence of short-range repulsion, which gives rise to an "exclusion zone" around each individual. This is equivalent to hard-core repulsion in particle systems. In order to measure the exclusion zone, we fitted the probability distribution of nearest neighbour distance to that of a hard-core system (Torquato 2002). We found that the size of the exclusion zone was very stable from flock to flock, and did not correlate with sparseness $r_1$ ($P$=0.56). The average over all flocks gives $r_h$=0.19m±0.02m SE. Therefore, the average minimum bird-to-bird distance ($2r_h$~0.38) is larger than starling's typical body-length (BL=0.2m), but is comparable to the typical wingspan (WS=0.4m).

### Nearest Neighbours Angular Distribution

Finally, we considered the angular distribution of the nearest neighbours. To do this, for each individual bird, we considered the vector to its nearest neighbour and measured the angle $\theta$ between this vector and the direction of motion of the flock. The distribution of $\cos(\theta)$ should be constant and equal to 0.5 for an isotropic system. For example, for a completely random arrangement of birds the neighbours can be found anywhere around the focal one. The probability of finding a neighbour at angle $\theta$ can be computed by assuming a constant bird density in space. If we choose the direction of motion as the polar axis, then this probability is simply given by $P_{random}(\theta)d\theta = 1/2\sin(\theta)d\theta$, where the factor $\sin(\theta)$ is the Jacobian from Cartesian to polar coordinates. Thus, even in a random isotropic system the distribution of $\theta$ is not constant. On the other hand, a simple change of variables gives $P_{random}(\cos(\theta))d\cos(\theta) = 1/2$, and the distribution of $\cos(\theta)$ is therefore constant for a random isotropic system (see Cavagna et al. 2008b for more details). It is therefore much more convenient to look at this distribution to distinguish anisotropic from isotropic arrangements of neighbours. We plotted the distribution of $\cos(\theta)$ for four different flocks (Fig. 8). In some flocks the distribution peaks at $\cos(\theta)$= 0 indicating that the nearest neighbour was more likely to be found on the plane perpendicular to the velocity (Fig. 8 c,d); in other cases (Fig. 8 a,b) there were two well defined peaks at intermediate angles indicating a more structured distribution in space. However, for all the analysed flocks, we found that there was a lack of neighbours along the direction of motion ($P(\cos(\theta))$ << 0.5 for $\cos(\theta) \sim \pm 1$).

## DISCUSSION

Our data constitute the first large-scale study of collective animal behaviour in a biological system. Some of these results (e.g. the exclusion zone or the density range) can be used as input parameters for existing models. Most of the results, however, should be used to refine and extend these models, to verify and assess their assumptions, and to identify the most appropriate theoretical framework. In addition, our results obviously have relevance for several general biological issues.

*Morphology*

Perhaps the most interesting morphological result is that flocks seem to have a characteristic shape, being thin in the direction of gravity and more extended perpendicular to it. The proportions of the flock are well-defined, with only weakly fluctuating aspect ratios, despite showing a wide range of sizes (Table 1). Our ability to conclude this stems entirely from the fact that we were able to analyse several groups with very different sizes (dimensions and number of birds). Non-spherical shapes have also been observed in fish schools, with the average proportions $I_1:I_2:I_3$ ranging from 1.0:1.7:2.1 in pilchards (Cullen et al. 1965) to 1.0:3.0:6.0 in saithes and 1:3:4 in herrings (Partridge et al. 1983). These values are comparable to those we found for starlings: 1.0:2.8:5.6. These studies, however, are limited in several respects: first, group size was limited to a few tens of individuals, so that it was not possible to check whether or not the proportions remained stable at different sizes; second, it is unclear to what extent the shape of the tank and the water depth influenced group morphology.

In our analysis, using a much larger statistical sample, we verified that, while the smallest dimension is strongly correlated with volume and number of individuals, the flock is organized in a way that keeps its proportions constant. This raises a number of interesting questions: do these proportions serve any function? Is it the individuals themselves or some external stimulus that keeps the group's proportions constant? If the individual birds that are responsible, how do they achieve this, starting from a purely local perception of the aggregation?

There may be a general reason why some specific proportions have been selected. For fish schools it has been suggested that a thin shape minimizes the chance of being seen by distant predators (Partridge et al. 1983). Alternatively, one could argue that this thin elongated shape is a direct consequence of the mechanisms leading to group formation and does not have any adaptive function. This correlation between elongation and velocity was suggested long ago on purely biological principles (Breder 1959; Radakov 1973) and was later observed in fish schools (Pitcher 1980), and has been demonstrated in numerical simulations of fish schools, which show that aggregations produced with simple local interactions acquire an elongated shape in the direction of the velocity (Kunz & Hemelrijk 2003). We did not, however, find any evidence of a this kind of pattern: the aspect ratios did not depend significantly on the absolute value of the velocity and there was no correlation between the longest axis and the velocity direction. This suggests that this feature may differ between birds and fish and that, in the case of bird flocks, the global shape of the groups may not be fully explained by the interactions between individuals.

One possible explanation for this difference is that aggregation dynamics between groups plays a role here. Recent models of aggregation-fragmentation (Okubo 1986; Bonabeau et al. 1999) have been used to predict successfully scaling in the size distribution of several species (from tuna fish to buffalo herds). These models may be helpful to investigate the origin of the shape of the groups. The rules of fission and fusion between groups, however, need to be generalized to this purpose. They were assumed to be isotropic in the past, while this may not be the case in reality. For example, studies on fish schools have shown that they become multilayered as the number of individuals increases (Partridge et al., 1980).

Gravity may also explain the thin shape that we observe: up/down space fluctuations are energetically more costly than lateral ones, favouring the stretching of the aggregation along the plane perpendicular to gravity. Finally, from an aerodynamic point of view, a simple model shows that there is a net decrease in drag when the flock extends further laterally than vertically (Higdon & Corrsin 1978).

*Orientation*

We found that all the flocks we analysed have a well-defined orientation in space: velocity and gravity are approximately perpendicular, while yaw and gravity are nearly parallel. In other words, flocks slide horizontally, flying almost parallel to the ground. This may seem odd to anyone who has observed flocks. The reason for this is that we typically watch from the ground, with our heads tilted-up, so we are unable to distinguish between closer and more distant birds. As a result, we project the flock onto our tilted plane of vision, thus losing perception of its orientation and relatively thin aspect. It is possible, however, that the flight orientation is different in more dramatic events than the ones we considered (as in the case of a predator's attack or for flocks at higher altitudes).

When looking for possible explanations of this behaviour, it is reasonable to expect gravity to play a strong role in the flock's relative orientation. Migratory flocks and flocks travelling from feeding to roosting sites exhibit level-flight parallel to the ground, and typically spread out horizontally rather than vertically (Major & Dill 1978). Several kinds of fish also form almost two-dimensional schools that swim parallel to the water surface (Cullen et al. 1965; Partridge et al. 1983). Gravity directly determines individual kinematics which occurs at a lower cost orthogonal to gravity: starlings perform steady level flight or intermittent undulating flight (temporarily gaining or losing altitude in correspondence of flapping and gliding periods) where, in any case, the optimal flight strategy corresponds to a mean flight path, which is level (Rayner et al. 2001). In other words, birds tend to fly, on average, parallel to the ground due to energetic considerations.

*Turns*

We observed that, during a turn, the flock's orientation changes with respect to velocity: if the flock was originally moving in the same direction as the intermediate axis, at the end of the turn it ends up oriented along the longest axis (see Fig. 4c). This explains why we did not find any statistically preferred orientation of flocks with respect to the direction of motion: flocks remain over the roost and continuously perform turns, so that their relative orientation with respect to velocity changes continuously.

Pomeroy and Heppner (1992) described the mechanism of turning in a group of 12 rock doves (*Columba livia*), and showed that birds turn in equal-radius paths, rather than parallel paths. For example, in a 90-degree left-turn, birds at the front of the flock end up on the right of it, while those on the left of the flock end up at the front of it (Fig. 4d). In this way what was the front-to-back axis before the turn becomes the side-to-side axis after the turn. The mechanism described by Pomeroy and Heppner (1992) is confirmed by our empirical observations here: if birds turn along equal-radius paths, it follows that, as we show, the velocity must rotate with respect to the flock's planar axes. Note that, unlike equal radius paths, parallel paths would require very different accelerations of the individuals during the turn, in order to retain the shape of the flock unaltered, which is what we observed. Energetic considerations would suggest this is unlikely to occur.

This turning mechanism leads to several interesting issues. From the point of view of the individual, group membership is advantageous for its anti-predator benefits. However, not all the positions in the aggregation are equivalent, and, under some circumstances, some positions may be more advantageous than others. For example, birds at the boundary of the flock typically suffer a higher predation risk. If the cost/benefit balance were negative for too many individuals or for too long, the group would eventually break up. Given that, in general, animal aggregations are rather stable, this implies that group structure and dynamics must allow for a systematic redistribution of risk among its members. Individuals must be able to move through the flock and exchange positions, while at the same time maintaining the integrity of the group. We found that the front, sides and back of the border were not stable regions of the flock. Rather, they continuously switched. This indicates that there were no long-lasting privileged positions along the border, so that the risk/benefit of any boundary location is periodically readjusted. It would be interesting to know whether there is an analogous turnover with respect to the border-centre direction. Future work on individual trajectories will allow us to answer this question.

### *Average Density and Nearest Neighbour Distance*

Our results on the internal structure of the groups showed that the density and the average nearest neighbour distance within flocks does not depend on the size of the group. This is contrary to observations of fish schools (Partridge et al. 1980) and the the results of theoretical models (Kunz & Hemelrijk 2003), where the average nearest neighbour distance was found to increase as the number of fishes in the group increased. However, in both of these studies, the average nearest neighbour distance was computed without taking into account the bias introduced by the border and, consequently, they do not give an accurate measure of group density. Individuals located on the border maintained an average nearest neighbour distance that was larger than that typical of internal individuals, simply because part of their surrounding volume was empty (see Cavagna et al. 2008b for a more detailed discussion of border effects). Given that the percentage of individuals on the border decreases as group size gets larger, ignoring the bias of the border means that the average nearest neighbour distance will inevitably decrease with group size, even if the density of groups remains the same at all sizes.

Another interesting result, shown clearly in Table 1 and Figure 5b, is that density varies considerably across flocks and there is not a well-defined typical value. Of course, this may be due to the limited number of flocks we analysed and a larger sample could reveal a distribution of densities that peaks at a characteristic value. Even if this were the case, however, it remains true that density fluctuations across flocks were very large. In addition, density did not depend on the size of the flock (contrary to observat small fish schools, Partridge 1980). This leaves open the question as to what exactly determines the density of a flock. Clearly, there are some relevant biological factors to consider. For example, the presence of predators (falcons) or disturbance actions (e.g. seagulls). When under attack, a flock exhibits fast expansions and contractions, indicating that strong perturbations have a direct effect on density. This cannot explain the variation we see here because the events we analysed did not involve predator attacks.

Many models of self-organized motion assume that the interaction between individuals depends on their mutual distance in space. In such cases, the average inter-individual distance (and therefore density) is determined by the nature of the relationship between inter-individual interactions with distance. An individual must be attracted to its neighbours over a certain threshold distancein order to remain part of the flock. At short distances, however, it is repelled, in order to avoid collisions. For a given distance, these forces of attraction and repulsion will compensate for each other and the neat force experienced by the individual will be null: this value becomes the average nearest neighbour distance of the model. In a recent work, however, (Ballerini et al. 2008), we have shown that interactions between starling individuals do not depend on their metric distance in space, but rather on their topological distance (i.e. whether they are first, second, third … neighbours). This means that there cannot be a well-defined distance in space where the force felt by one bird is null and therefore the global density is not simply related to microscopic interactions.

### Density Variations Within the Aggregation

All flocks exhibited a density gradient: density was higher at the border than in the centre. This result is quite surprising, given that some models of collective animal behaviour predict exactly the opposite (see e.g. Kunz & Hemelrijk 2003). The explanation for this is possibly to be found in the anti-predator response of the aggregations.

The evolutionary motivation for grouping has been associated traditionally with its anti-predator function (Pitcher & Parrish 1993; Vine 1971; Parrish 1992). Belonging to a group of similar individuals decreases the probability of being caught (*dilution effect*). Moreover, moving together also reduces the ability of predator to focus on a specific individual and capture it. The response to predators is likely to optimize this *confusion effect*, as can be seen in the very effective escape manoeuvres displayed by starling flocks under attack. The high density borders that we observed may represent a feature that enhances such anti-predatory tactics, creating a 'wall' effect to increase the predator's confusion. However, we do not at present understand how the density gradient is produced in terms of inter individual dynamics within the group. Reconstruction of the individual dynamical trajectories, which is currently underway, will help to clarify this point.

Finally, some numerical models of fish schools reveal that density is higher at the front of the group (Kunz & Hemelrijk 2003; Hemelrijk & Kunz 2004; Hemelrijk & Hildenbrandt 2007), as also observed in natural shoals (Baumann et al. 1997). In our analysis however we did not find a clear and general correlation between the densest part of the flock and the direction of motion.

*Nearest Neighbours Radial Distribution and Exclusion Zone*

An important result concerned the presence of a well-defined exclusion zone around individual birds, which was very stable from flock to flock and whose diameter was comparable to the average wingspan. The finding that birds do not come too close one to the other is highly intuitive, given the need to avoid collisions, and is consistent with one of the main assumptions of numerical models. Models for flocking, which use the size of the exclusion zone as an input parameter, can now be fed with our empirical value, in order to increase their accuracy.

We stress that the value of the exclusion zone (~0.38m) does not depend on density, and that it provides a metric scale characteristic of flocking behaviour. As already mentioned, a second study on the same data set (Ballerini et al. 2008) shows that the interaction between birds possesse a crucial topological character: each individual interacts with up to 6-7 neighbours, irrespective of their distance. It seems, therefore, that the interaction between birds can be understood at two levels: metric at short scales, and topological at larger ones. In other words, when a neighbour is too close, a bird takes into account its physical distance and tries to keep it outside the exclusion zone; for more distant birds, however, the actual distances do not matter and the bird interacts up to its seventh neighbour, wherever it is located in space. The reasons for such behaviour are inherent in the very nature of flocking and aerial display. On the one hand, individuals in a group must avoid collisions and control the mutual dynamics at short distances. On the other hand, the flock undergoes large density variations that can modify dramatically the distances of its nearest neighbours: to keep the same degree of cohesion, indispensable during predator attacks, each bird must keep track of the same number of neighbours, even if their distance changes.

The analysis of nearest-neighbours distribution using a hard-sphere model also revealed another quantitatively important feature: flocks are not as compact as they may appear to an observer. For hard-spheres, the degree of compactness of the system can be classified through the so-called packing fraction: $\phi = 4/3 \, \pi \rho \, r_h^3 = v \, N / V$, where $\rho$ is the density, $r_h$ the size of the hard-core (the exclusion zone), $v$ is the volume of each sphere, $N$ is the total number of spheres, and $V$ is the volume of the aggregation. The packing fraction is simply the ratio between the total volume occupied by the spheres and the volume of the aggregation. It depends only on the hard-core value and on the density and, by definition, it is a number between 0 and 1. Small values of $\phi$ correspond to very diluted systems (gas-like), while large values correspond to compact ones (liquid or crystals). To give some quantitative reference values, compact crystalline arrangements of spheres correspond to $\phi > 0.490$, while the value for the most compact possible arrangement is $\phi \sim 0.79$ (the so-called random close packing, Torquato 2002). All the flocks we analysed showed packing fractions smaller than 0.012. This is a very small value indeed. In hard-sphere language, flocks are extremely sparse systems, with much lower densities than a crystalline arrangement. This sparseness is not evident to the eye, since the two-dimensional projection we perceive looks much denser. This sparseness makes good biological sense because, if the birds were too close one to the other, exchange of positions and diffusion inside the flock would be difficult to achieve. Further analysis on individual birds trajectories will allow us to investigate quantitatively the extent to which the individual freedom to move is related to the overall flock density.

Despite being such sparse systems, flocks exhibit a non-trivial structure in space, which is well described by the angular distribution of nearest neighbours. Neighbours are less likely to be found along the direction of motion, and instead they concentrate laterally. While the comparison with particle systems can be extremely useful from a methodological point of view, one always has to keep in mind that animal groups are very different from particle systems. A hard-sphere system with the same hard-core and density values as those we found here would be a gas without any structure, while flocks exhibit a strongly anisotropic structure. Reproducing such features will be an on-going challenge for models and theories.

*Nearest Neighbours Angular Distribution*

A similar spatial anisotropy to the one found here for starlings has been reported in fish schools (Cullen 1980), suggesting this is a typical feature of collective behaviour in both birds and fish. An important question, therefore, is to understand the origin of this anisotropy. One possibility is that it is simply an effect of the existence of a preferential direction of motion. However, the simplest numerical models of self-organized motion, which assume isotropic interactions between individuals, give a non-zero velocity for the aggregation, but fail to reproduce the angular anisotropy. This suggests that the anisotropy in the nearest neighbour's distribution is an explicit consequence of the anisotropic character of the interaction itself, which models should incorporate into their assumptions if they want to reproduce the observed behaviour.

There are various reasons why interactions between individuals should be anisotropic. First of all, we note that vision itself has an anisotropic nature in both birds and fishes. In particular, starlings have lateral visual axes and a blind rear sector (Martin 1986). Thus, if vision is the main mechanism by which interactions are controlled, then the very structure of the eye may be responsible for the lack of nearest neighbours in the front-rear direction (Heppner 1974; Badgerow 1988; Speakman & Banks 1998). An alternative idea is that the mutual position chosen by the animals is the one that maximizes the sensitivity to changes of heading and speed of their neighbours (Dill et al. 1997). According to this hypothesis, even though the interaction is still vision-based, it is an optimization mechanism to determine the anisotropy of neighbours. Another possibility is that individuals try to keep a larger distance between themselves and the individuals in front of them, to avoid the risk of collisions in case of sudden changes of velocities. Finally, a radically different claim is that anisotropic structures in both bird and fish aggregations saves energy thanks to aerodynamic (or hydrodynamic) advantages (Lissaman & Shollenberger 1970; Weihs, 1973; Hummel 1995). The energy-saving principle has been challenged, however, for both birds (Badgerow & Hainsworth 1981) and fish (Partridge & Pitcher 1979). More importantly, at least for aerial display, the fact that inter-individual interactions depend on the order of neighbour, rather than on their distance, rules out any aerodynamic arguments, because these imply a strong dependence on the metric distance (as discussed in Ballerini et al. 2008).

In this paper we presented large-scale data on starling flocks during aerial display. Our data were obtained in field observations of large, naturally occurring groups. This is, we believe, a very important feature, which distinguishes our research from previous ones. Experiments in the laboratory may condition some important features of the groups, such as their shape and dynamics, due to the confined space. Groups of small size, both in the laboratory and in the field, do not allow a reliable statistical analysis due to border effects. Thanks to our experimental procedure, we avoided these two problems and produced, for the first time, unbiased data on very large groups in the field.

We investigated the main features of the flocks, shape, movement, density and structure, and characterized them as emergent attributes of the grouping phenomenon. Our data provide with a new experimental benchmark for testing and improving theoretical models of self-organized motion. In this light, we discussed some of our results in connection with the known predictions of existing models. We hope that our analysis will help to clarify what are the fundamental microscopic mechanisms leading to collective behaviour in animal groups, and how appropriate behavioural rules for the individuals can determine specific features of the aggregation at the group level.

## Acknowledgements

We thank E. Alleva, F. Bartumeus, C. Carere, G. Cavagna, I. Couzin, D. Grunbaum, C. Hemelrijk, H. Hildelbrandt, D.Santucci, D. Stoyan for interesting discussions. We particularly thank A. Cimarelli and F. Stefanini for technical help and several discussions.

## Table 1.  Global quantitative properties

| Event | Number of birds | Volume $(m^3)$ | Density $\rho$ $(m^{-3})$ | Nnd $r_1$ (m) | Velocity $(ms^{-1})$ | Concavity | Balance shift | Thickness $l_1$ (m) | Aspect ratios $l_2/l_1$ | $l_3/l_1$ | Orientation Parameters $\mathbf{I_1}\cdot\mathbf{G}$ | $\mathbf{V}\cdot\mathbf{G}$ | $\mathbf{V}\cdot\mathbf{I_1}$ |
|---|---|---|---|---|---|---|---|---|---|---|---|---|---|
| 32-06 | 781 | 930 | 0.8 | 0.68 | 9.6 | 0.03 | 0.08 | 5.33 | 2.97 | 4.02 | 0.89 | 0.06 | 0.20 |
| 28-10 | 1246 | 1840 | 0.54 | 0.73 | 11.1 | 0.34 | -0.06 | 5.29 | 3.44 | 6.93 | 0.80 | 0.09 | 0.41 |
| 25-11 | 1168 | 2340 | 0.38 | 0.79 | 8.8 | 0.37 | -0.1 | 8.31 | 1.90 | 5.46 | 0.92 | 0.12 | 0.14 |
| 25-10 | 834 | 2057 | 0.34 | 0.87 | 12.0 | 0.05 | 0.0 | 6.73 | 2.65 | 4.98 | 0.99 | 0.18 | 0.18 |
| 21-06 | 617 | 2407 | 0.24 | 1.00 | 11.2 | 0.04 | 0.0 | 7.23 | 2.56 | 4.53 | 0.96 | 0.09 | 0.11 |
| 29-03 | 448 | 2552 | 0.13 | 1.09 | 10.1 | 0.20 | 0.0 | 6.21 | 3.58 | 5.96 | 0.97 | 0.27 | 0.06 |
| 25-08 | 1360 | 12646 | 0.09 | 1.25 | 11.9 | 0.19 | 0.16 | 11.92 | 3.32 | 5.12 | 0.95 | 0.14 | 0.12 |
| 17-06 | 534 | 5465 | 0.08 | 1.30 | 9.1 | 0.18 | 0.5 | 9.12 | 2.76 | 6.94 | 0.91 | 0.09 | 0.32 |
| 16-05 | 2631 | 28128 | 0.06 | 1.31 | 15.2 | 0.15 | 0.0 | 17.14 | 2.46 | 8.36 | 0.90 | 0.19 | 0.25 |
| 31-01 | 1856 | 33487 | 0.04 | 1.51 | 6.9 | 0.24 | 0.17 | 19.00 | 2.44 | 4.07 | 0.95 | 0,09 | 0.13 |

Flocking events are labelled according to session number (each day of data-taking corresponding to a session) and position within the session (in temporal order). Each quantity is averaged over the different shots of the event. Events are ordered by increasing values of the average nearest neighbour distance (Nnd), $r_1$. The density is computed as the number of internal birds divided by the volume of the flock, as defined with the $\alpha$-shape. Velocity refers to the centre of mass (the average position of all birds in the flock). Concavity is defined as the relative volume difference between actual and convex border of the flock. The balance shift is defined as the relative difference along the direction of motion between the position of the centre of mass and the geometrical centre of the flock: positive values indicate that the density is larger on the front.  The thickness $l_1$, the two larger dimensions $l_2$ and $l_3$, and the corresponding aspect ratios, are defined as described in the Methods section. The three last columns report the scalar products (absolute values) between yaw (the axis relative to the shortest dimension), gravity and velocity.

**FIGURES**

Figure 1

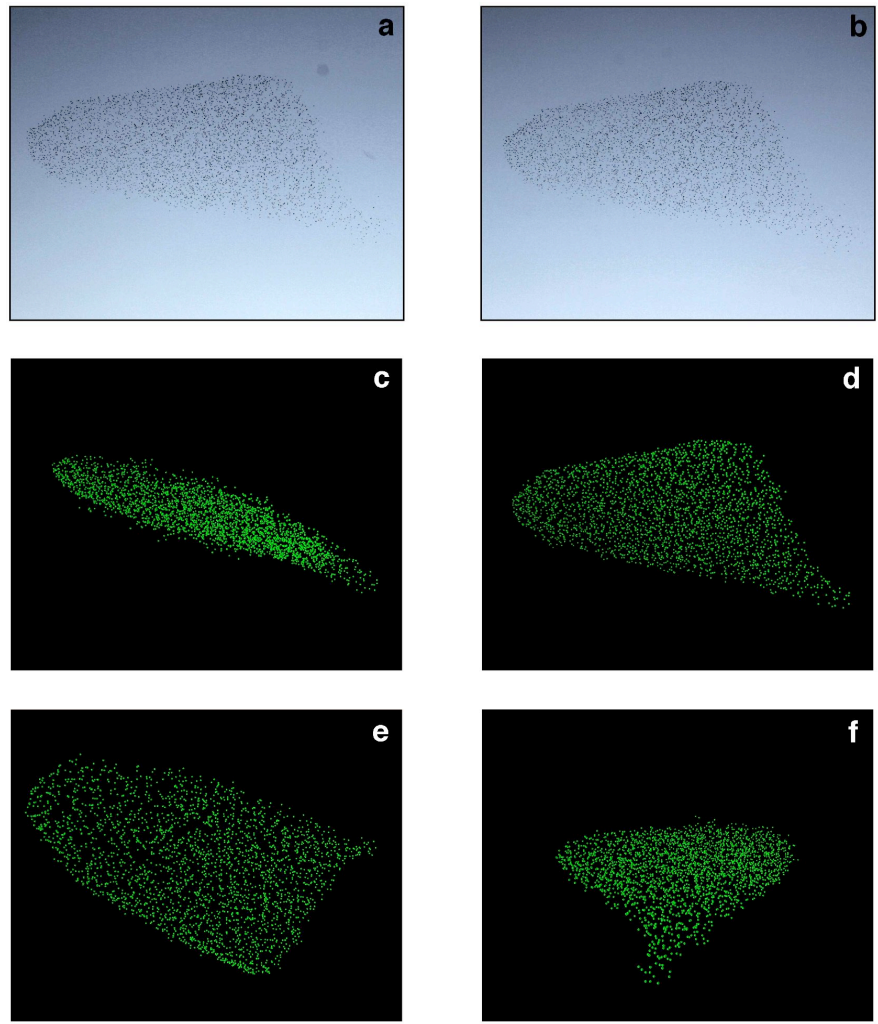

## Figure 1. A typical flock and its 3D reconstruction

This flock belongs to event 16-05, and consists of 2630 starlings, flying at approximately 240 m from the cameras. The cameras tilt-up was 40%. **a,b**, Left and right photographs of the stereo pair, taken at the same instant of time, but 25 meters apart. To perform the 3D reconstruction, each bird's image on the left photo must be matched to its corresponding image on the right photo. **c,d,e,f**, 3D reconstruction of the flock in the reference frame of the right camera, under 4 different points of view. The thinness of the flock is evident. Panel **d** shows the reconstructed flock with the same perspective as the right photograph (**b**).

Figure 2

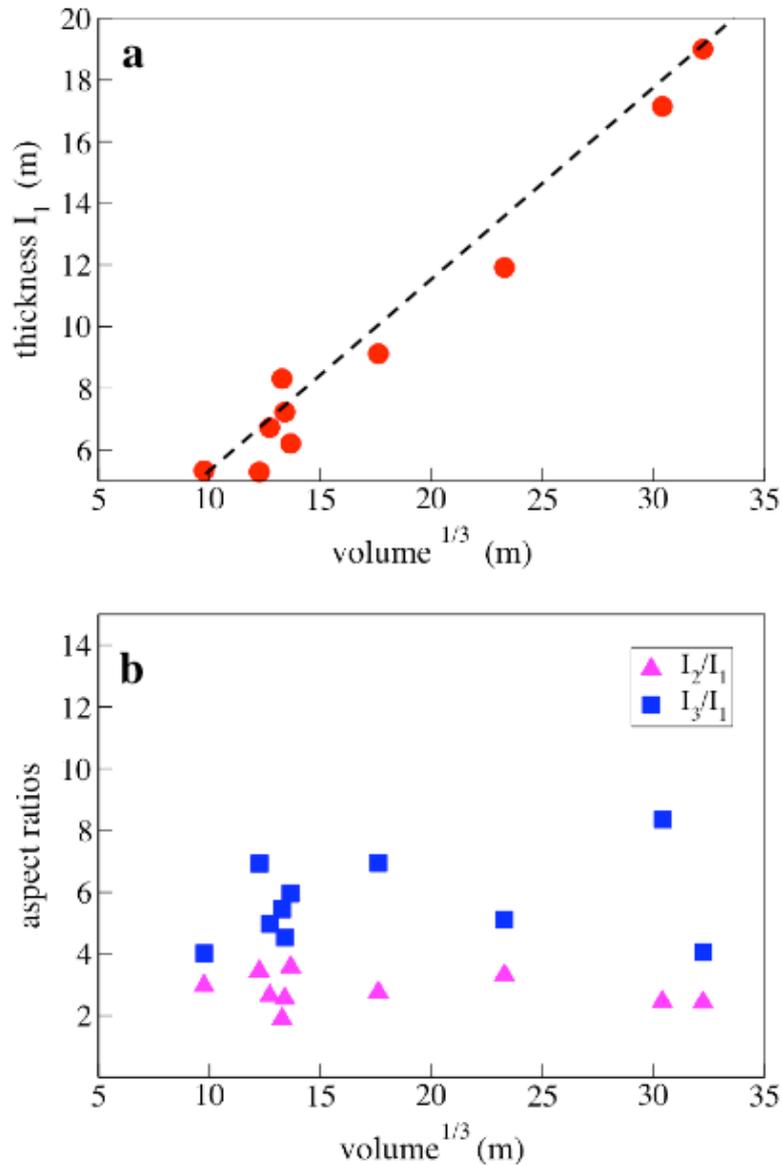

**Figure 2. Dimensions and aspect ratios**

**a.** The flock's thickness $l_1$ is plotted against $V^{1/3}$, where $V$ is the volume, for all the ten events we analysed. The dash-dotted line represents a linear fit of the data, showing a clear linear correlation between these two quantities (Pearson, $n$=10, $R^2$=0.97, $P$=1.6×10$^{-7}$). **b.** The aspect ratios, i.e. the ratios of the two longer dimensions to the shortest dimension, $l_2/l_1$ and $l_3/l_1$, are rather stable and show no significant correlation with volume ($P$=0.51 and $P$=0.54, respectively).

**Figure 3**

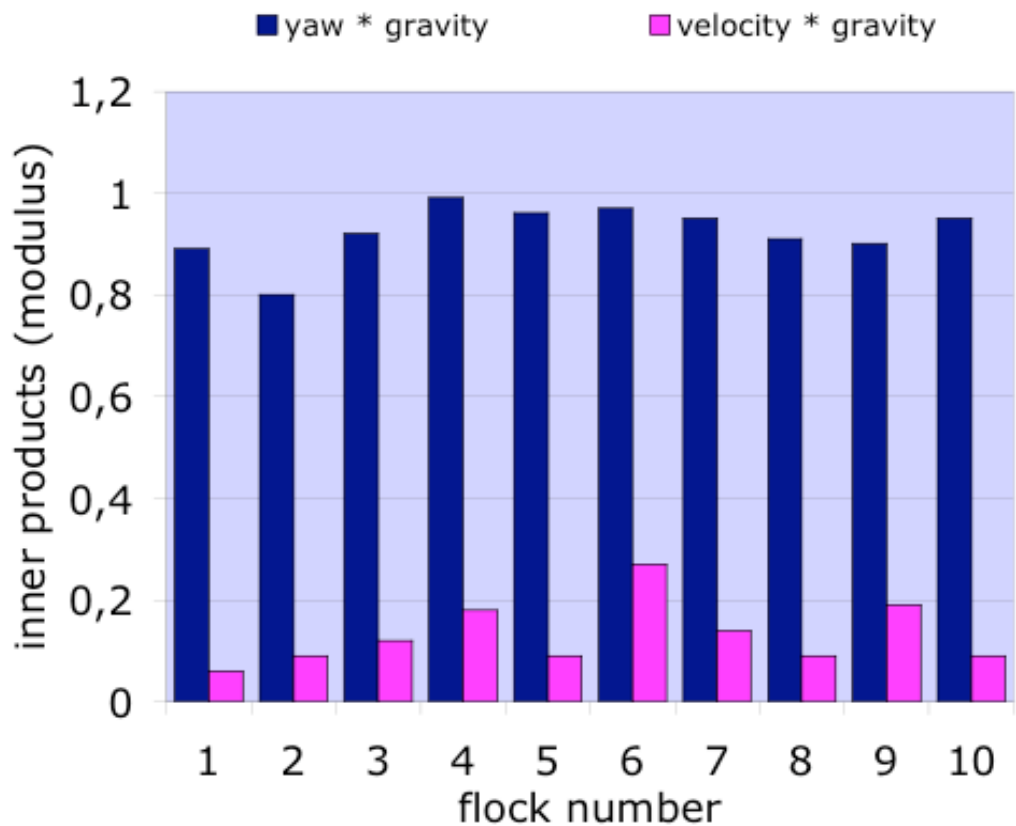

**Figure 3. Orientation**

The flock's orientation in space can be described by the mutual orientation of yaw ($l_1$), velocity (**V**) and gravity (**G**). The inner product (in modulus) between yaw and gravity ( $|l_1 \cdot G|$ ) is close to one in most flocks. The inner product (in modulus) between velocity and gravity ( $|V \cdot G|$ ), on the other hand, is small in most events. These results show that most of the times flocks slide horizontally, flying approximately parallel to the ground.

**Figure 4**

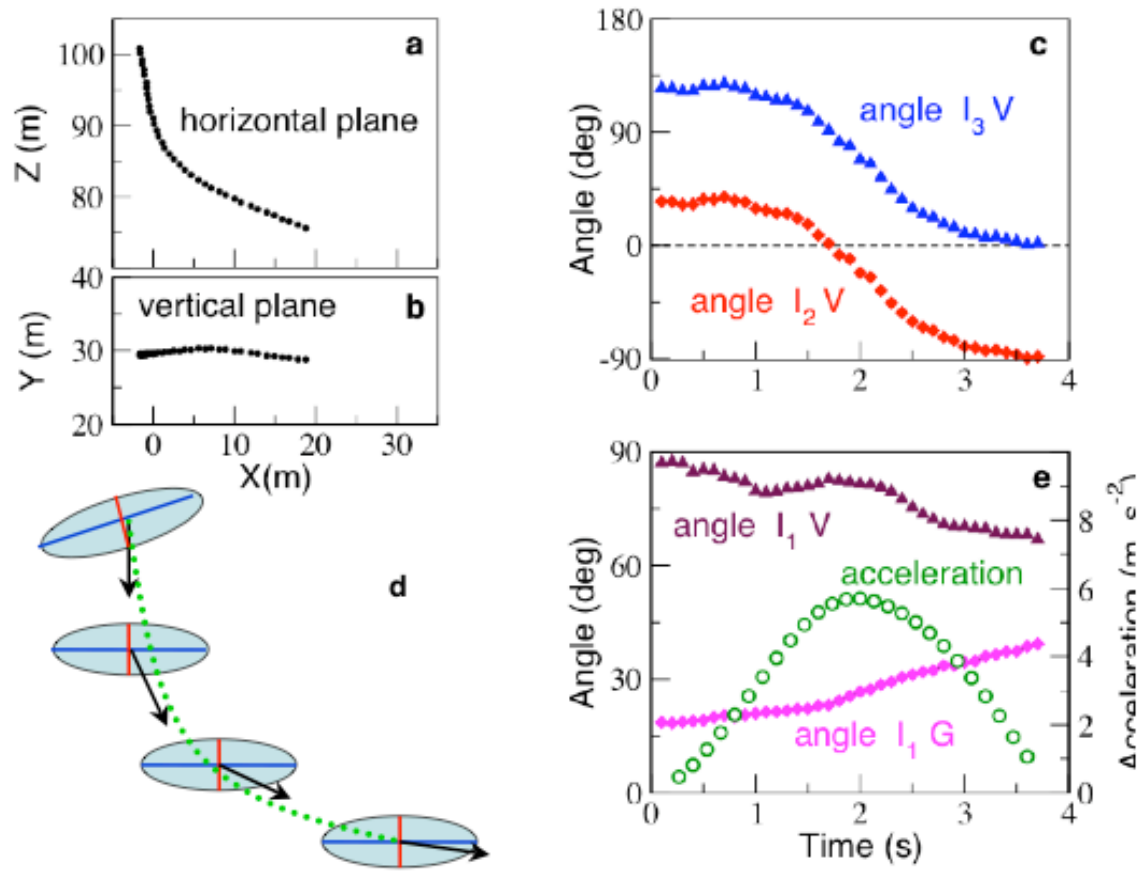

## Figure 4. A turning flock

Event 32-06 lasts 4 seconds before the flock exits the camera's field of view. **a,** The trajectory of the flock projected on the horizontal plane (orthogonal to gravity), clearly shows a turn (direction of motion is left to right). Each point represents the (X,Z) position of the centre of mass of the flock, at time intervals of 0.1s. **b,** The same trajectory, when projected on the vertical plane (X,Y) (with Y parallel to gravity) looks very thin, showing that the flock flies parallel to the ground. **c,** The angles between the planar orthogonal axes $I_2$ and $I_3$ of the flock, and the projection of the velocity **V** on the plane vs. time. The velocity rotates with respect to $I_2$ and $I_3$: at the beginning of the turn the velocity is approximately perpendicular to the longest axis $I_3$, whereas **V** and $I_3$ are parallel at the end of the turn. **d,** Schematic representation of the flock's turn on the horizontal plane; $I_2$ (red), $I_3$ (blue) and **V** (black) have mutual orientation as in the real event. Apart from a slight rotation at early times, the absolute orientation of the flock remains constant throughout the turn. **e,** The yaw axis $I_1$, orthogonal to the flock's plane, is almost parallel to gravity and perpendicular to velocity before the turn. As the turn approaches, the angle between $I_1$ and **G** increases, whereas the angle between $I_1$ and **V** decreases. Hence, the yaw axis tilts in the direction of the turn, and the flock's plane is no longer parallel to the ground. The centripetal acceleration (in m/s$^2$) has been computed from a Gaussian SP-line interpolation of the discrete trajectory. As expected, the acceleration shows a peak associated to the turn.

Figure 5

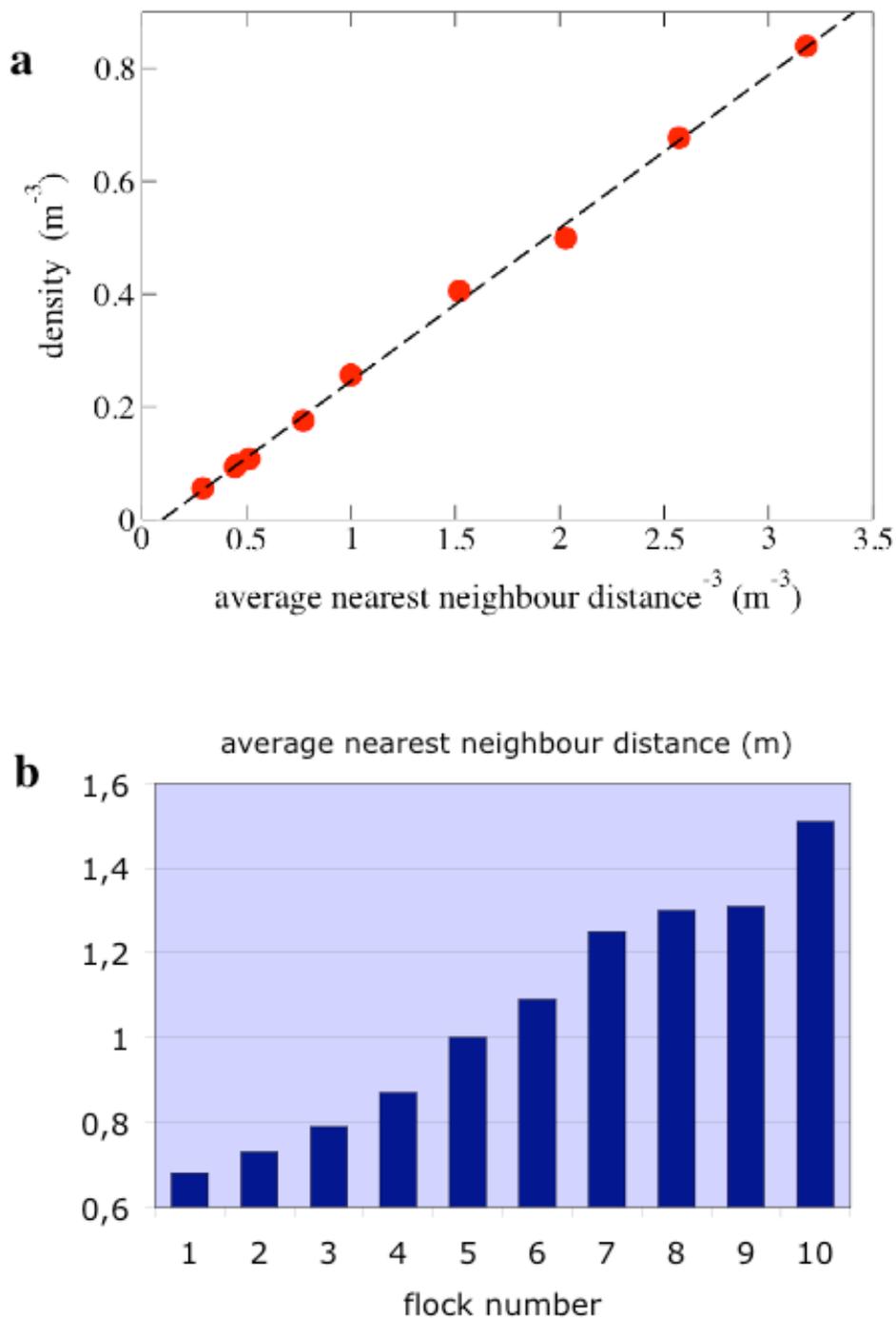

**Figure 5. Nearest neighbour average distance**

**a.** The density ρ is proportional to $r_1^{-3}$ (Pearson correlation test: $n$=10, $R^2$=0.8, $P$=0.0004) and we can then use $r_1$ as a convenient measure of a flock's sparseness. **b.** The sparseness $r_1$ varies significantly from flock to flock, ranging from 0.7m to 1.5m, in the analysed events. Average body length and wingspan of starlings are respectively BL=0.2m and WS=0.4m, and thus in the densest flocks $r_1$~3.5BL and $r_1$~1.7WS.

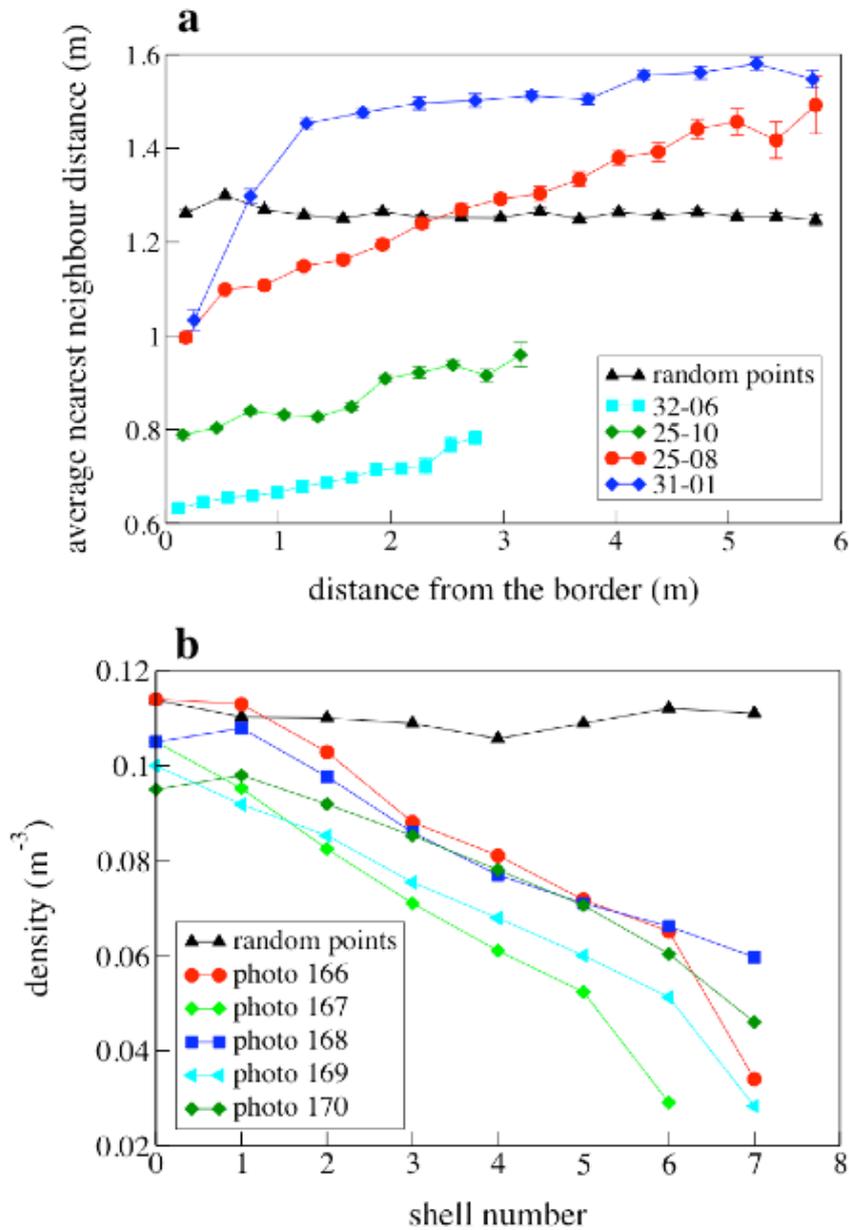

Figure 6

**Figure 6. Density gradient**

Flocks exhibit a clear density gradient from the border to the centre. **a.** Average nearest neighbour distance as a function of the distance from the border, for 4 flocking events. Each curve is obtained by averaging over the consecutive shots of the corresponding event (error bars indicate standard errors due to the corresponding fluctuations). In all the flocks the average nearest neighbour distance increases when going from the border to the centre, indicating that flocks are denser at the border. We also report the behaviour of a Poisson ensemble of random points of density and dimensions comparable to event 25-08: for a homogeneous system the average nearest neighbour distance is constant and does not depend on the distance from the border. **b.** Density as a function of the shell number. To compute the density, the birds in a shell of length $\delta=0.2$ from the border are erased, then the border of the reduced set of birds is re-computed, and finally its density obtained as the ratio of the number of internal birds by the reduced volume. The procedure is repeated until no birds are left in the reduced set. The curves correspond to single consecutive shots in event 25-08, where the density gradient is more evident and the number of birds is large enough to have a single-sample good statistics. To compare with the random homogeneous case, and to be sure that small volume effects do not bias the decrease in density, we generated a set of random points with the same border, volume and internal density as photo 166 of flock 25-08. In the other events the density gradient is qualitatively similar.

## Figure 7

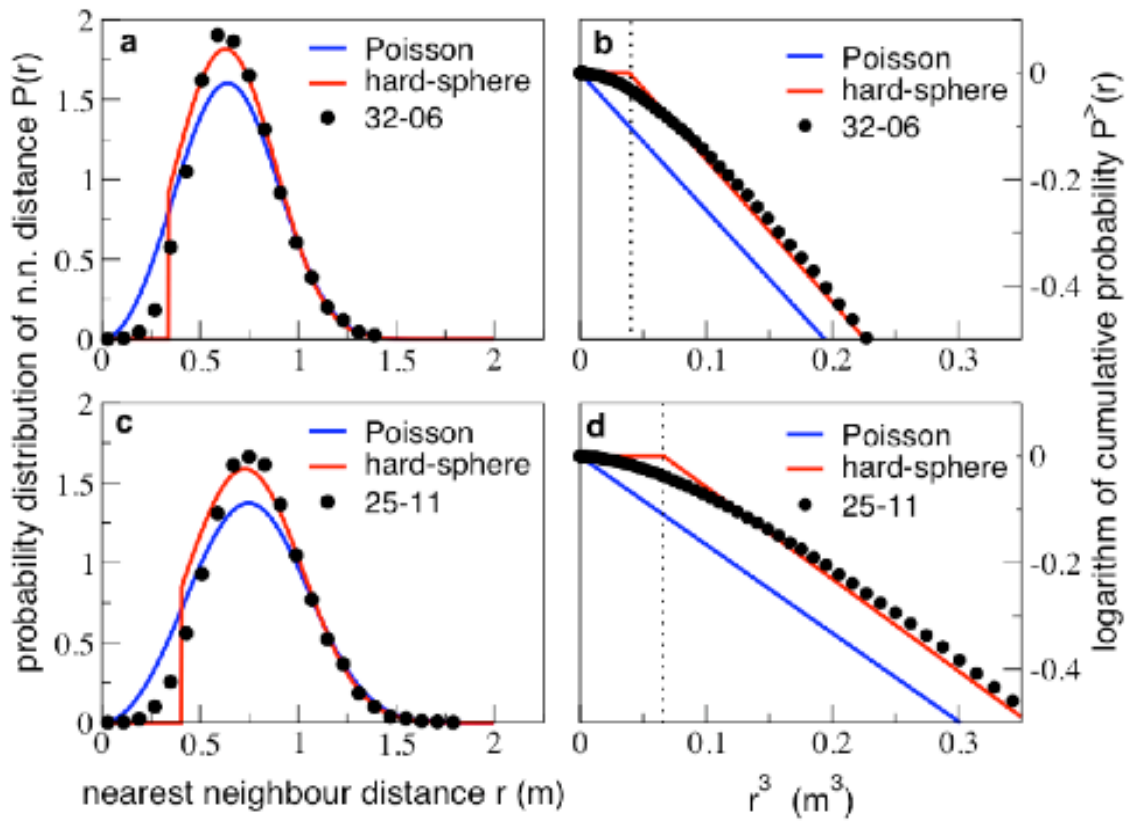

### Figure 7. Nearest neighbour distance distribution and hard core

**a,** Event 32-06: probability distribution $P(r)$ of the nearest neighbour distance. The average n.n. distance $r_1$ is the average of $r$ with this distribution. Circles are the data. The red line is the hard-sphere fit, the blue line is the Poisson fit. Compared to Poisson, data show a drop of the probability at small values of $r$. Birds (on average) tend not to get closer than a certain minimum distance**.** In this sense, we say that birds have an effective hard-core $r_h$. As a consequence, the n.n. probability distribution of a hard-sphere system, which is zero for $r < 2r_h$, is a better fit of the data. **b,** The presence of a hard-core is particularly evident when one plots the logarithm of the cumulative probability $P^>(r)$ (the probability that the n.n. distance is larger than $r$) as a function of $r^3$. Circles are the data. The red line is the hard-spheres fit, the blue line is the Poisson case. The hard-sphere cumulative probability is equal to 1 for $r < 2r_h$, where $r_h$ is the hard-core. Thus, compared to Poisson, the hard-sphere curve has a shift due to the presence of the hard-core. This shift is clearly present also in the real data. For event 32-06 the hard-sphere fit gives $r_h$=0.17m (the value of $2r_h$ is marked by the vertical dotted line). **c,d,** Event 25-11, same symbols as in **a,b.** Hard-core $r_h$=0.20m .

Figure 8

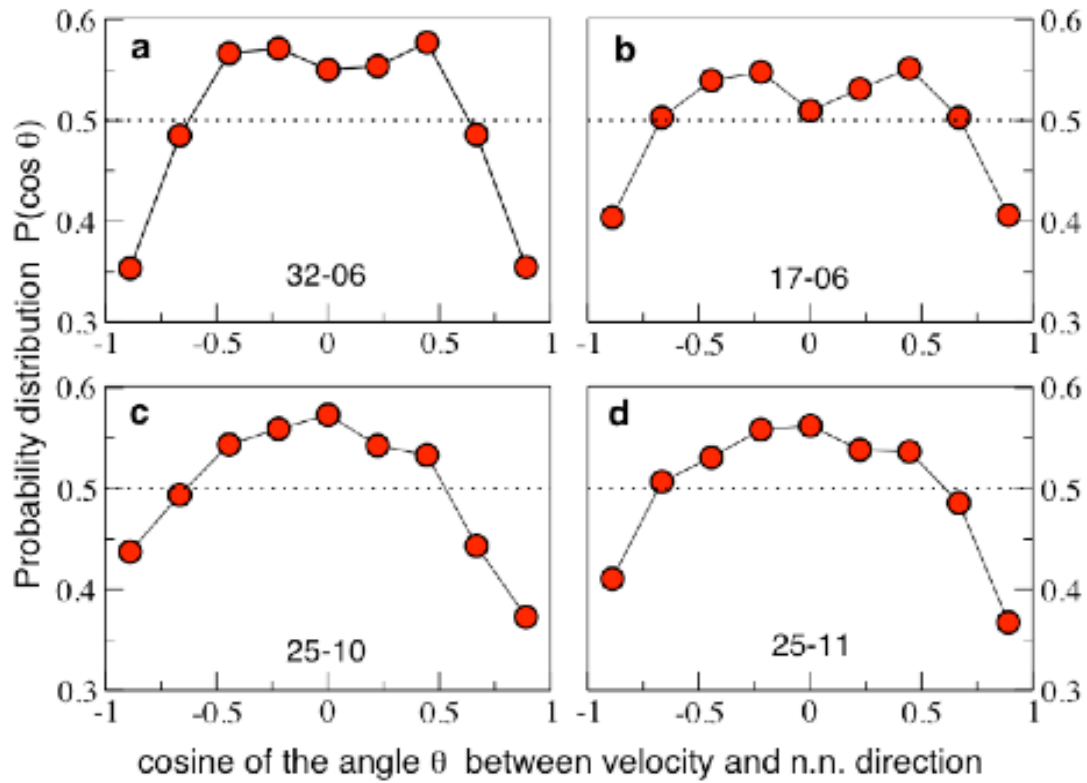

**Figure 8. Nearest neighbour angular distribution**

**a-d,** Probability distribution of the cosine of the angle θ between the global flock's velocity and the direction of the nearest neighbour, for events 32-06, 17-06, 25-10, and 25-11. cos(θ)=+1 corresponds to the front of the reference bird, whereas cos(θ)=-1 corresponds to the back. For an isotropic arrangement of points there is no correlation between velocity and position of the nearest neighbour, therefore all directions are equally probable, and this probability distribution is constant and equal to 1/2 (provided that boundary effects are appropriately accounted for). All 10 flocks we analysed show a significant drop of the probability at cos(θ)=±1, indicating that the arrangement is not isotropic and that the nearest neighbour of each bird is less likely to be found around the direction of motion.